\documentclass[twocolumn,showpacs,aps,prl]{revtex4-1}

\newcommand{\beq}{\begin{eqnarray}}
\newcommand{\eeq}{\end{eqnarray}}
\newcommand{\beqq}{\begin{eqnarray*}}
\newcommand{\eeqq}{\end{eqnarray*}}

\usepackage{graphicx}
\usepackage{dcolumn}
\usepackage{bm}
\usepackage{color}

\begin{document}

\begin{titlepage}

\title{Crystal Field Effect Induced Topological Crystalline Insulators In Monolayer IV-VI Semiconductors}

\author{Junwei Liu,$^1$ Xiaofeng Qian$^2$, and Liang Fu$^1$}

\address{$^1$Department of Physics, Massachusetts Institute of Technology, Cambridge, MA 02139;\\$^2$Department of Materials Science and Engineering, Dwight Look College of Engineering, Texas A{\rm \&}M University, College Station, TX 77843.}

\begin{abstract}
Two-dimensional (2D) topological crystalline insulators (TCIs) were recently predicted in thin films of the SnTe class of IV-VI semiconductors, which can host metallic edge states protected by mirror symmetry.  As thickness decreases, quantum confinement effect will increase and surpass the inverted gap below a critical thickness, turning TCIs into normal insulators.  Surprisingly, based on first-principles calculations, here we demonstrate that (001) monolayers of rocksalt IV-VI semiconductors XY (X=Ge, Sn, Pb and Y= S, Se, Te) are 2D TCIs with the fundamental band gap as large as 260 meV in monolayer PbTe, providing a materials platform for realizing two-dimensional Dirac fermion systems with tunable band gap. This unexpected nontrivial topological phase stems from the strong {\it crystal field effect} in the monolayer, which lifts the degeneracy between $p_{x,y}$ and $p_z$ orbitals and leads to band inversion between cation $p_z$ and anion $p_{x,y}$ orbitals. This crystal field effect induced topological phase offers a new strategy to find and design other atomically thin 2D topological materials.

\end{abstract}

\maketitle

\draft

\vspace{2mm}

\end{titlepage}

%\section{introduction}
Topological insulators (TIs) have attracted extensive attention owing to their fundamental theoretical interests and potential applications \cite{kanehasan, zhangreview, moore}. The interplay between crystallography and electronic band topology has further given birth to a new type of topological phases, termed topological crystalline insulators (TCIs) \cite{fu, fu_rev}. The first example of TCIs was recently predicted in SnTe class of IV-VI semiconductors \cite{hsieh}. They host even number of Dirac cones on their (001) and (111) surface \cite{hsieh,jw_kp,hsin,safaei}, which were subsequently observed \cite{ando,poland,hasan,yan,Polley,Tanaka}. The metallic boundary states of TCIs are protected by crystal symmetry rather than time-reversal in TIs, and therefore these states can acquire a band gap under perturbations that break the crystal symmetry by either distortion \cite{hsieh, vidya, vidya2, serbyn} or electric field \cite{jw_tf}. In addition to fundamental interests, TCIs may realize novel device applications \cite{ezawa, fang, fzhang, qian}. Recent studies have further predicted/proposed a wide class of new TCI materials in antiperovskites \cite{A3BO}, pyrochlore iridates \cite{fiete}, multi-layer graphene \cite{kindermann} and heavy-fermion compounds \cite{dai,sun}.

More recently, based on tight-binding (TB) calculations \cite{lent}, we predicted that two-dimensional (2D) TCIs exist in (001) multi-layer thin films of SnTe and Pb$_{1-x}$Sn$_{x}$Se(Te) \cite{jw_tf}. It was found that thin films thicker than a critical thickness are topological nontrivial with metallic edge states protected by mirror symmetry $z\rightarrow -z$ ($z$ is normal to the film). As thickness of TCI thin films decreases, the quantum confinement effect will increase and eventually overcome the inverted gap, turning these 2D TCIs into normal insulators \cite{jw_tf}. On the other hand, it is known that when materials become {\it atomically} thin, the reduced dimensionality may lead to physical properties dramatically different from their 3D counterparts, as found in graphene \cite{geim1,geim2}, monolayer MoS$_2$ \cite{Hone, Qian}, monolayer $\alpha$-Sn \cite{xuyong} and FeSe \cite{Xue}.
At the same time, there are great interests in the exploration of nanostructure of TCIs in experiments\cite{Safdar, taskin, Judy2, hguo, yan, Assaf, Sasaki}. These motivates us to study IV-VI semiconductors in the monolayer form, whose electronic structures may not be captured by the bulk TB parameters \cite{lent} used in previous works.

Here, using first-principles calculations, we systematically studied the electronic structures of (001) monolayer IV-VI semiconductors XY (X= Ge, Sn, Pb and Y= S, Se, Te) in the  rocksalt structure. Remarkably, we found all of them are 2D TCIs and possess metallic edge states protected by $z \rightarrow -z$ mirror symmetry, contrary to the expectation from the quantum confinement effect. This unexpected nontrivial topological phase originates from the band inversion due to the strong crystal field effect on $p$ orbitals in the monolayer, which significantly lowers (raises) the onsite energy of cation (anion) $p_z$ orbital, thereby facilitating or even driving the band inversion towards the TCI phase. It is worth noting that this novel mechanism does not rely on the electronic topology of the 3D bulk states: for example, despite that PbTe is topologically trivial in the bulk form, monolayer PbTe is a 2D TCI with a large fundamental gap of 260 meV. Therefore our work reveals a unique advantage of atomically thin  2D materials in realizing 2D topological phases.

\section{RESULTS AND DISCUSSION}

\begin{figure}[tbp]
\includegraphics[width=9cm]{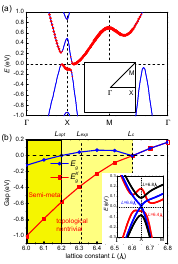}
\caption{(a) Band structure of monolayer SnTe, where the size of red dots indicates the weight of Sn $p_z$ orbital; the inset is the 2D Billouin zone of (001) thin films; (b) Band gap at $X$ point ($E_g^{X}$) and fundamental gap ($E_g$) of monolayer SnTe as a function of lattice constant ($L$). As $L$ increases, monolayer SnTe undergoes a topological phase transition from $|N_M|=2$ to $N_M=0$. When $L>6.2$~{\rm \AA}, it is a fully gapped semiconductor, otherwise, it is a semi-metal;  The inset is the band structure evolution around the transition point (red:$L$=6.4~\AA; blue: $L$=6.6~\AA; black: $L$=6.8~\AA).}
\label{bulk}
\end{figure}

\begin{figure}[tbp]
\includegraphics[width=7cm]{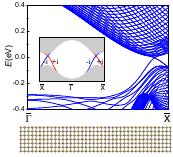}
\caption{Edge states of monolayer SnTe along [001] direction with lattice constant $L=6.4$~{\rm \AA} (top panel) and the schematic (001) ribbon model (bottom panel). The inset is schematic edge states for the whole Brillouin zone.}
\label{edge}
\end{figure}

The band structure plot displayed in Fig.~\ref{bulk}(a) shows that monolayer SnTe is a semiconductor with a small indirect gap located around the $X$ point. Moreover, the nearby $W$-shape conduction band and $M$-shape valence band provide a clear sign of band inversion at $X$ point with a large inverted gap (denoted by ${E_g^{X}}$) of about $-0.6$ eV at its optimized equilibrium lattice constant of $L_{\rm opt}=6.2$~{\rm \AA}. Negative sign of ${E_g^{X}}$ indicates an {\it inverted} gap. To confirm this band inversion, we performed first-principles calculations of monolayer SnTe under a series of biaxial elastic strain \cite{hsieh,jw_strain, qian}.  As shown in Fig.~\ref{bulk}(b), when lattice constant ($L$) increases, ${E_g^{X}}$ monotonically enlarges and reaches zero at a critical lattice constant of $L_{\rm c}=6.6$~{\rm \AA}. Beyond  $L_{\rm c}$, ${E_g^{X}}$ becomes positive, indicating a trivial topological phase. The monotonical change of ${E_g^{X}}$ is very similar to those in Bi2Se3-type topological insulator~\cite{jw_strain}, which comes from the different parities of bottom conduction band and top valence band at X points: one corresponds to bonding states and the other corresponds to anti-bonding states, whose energy changes oppositely under the biaxial strain. However, the fundamental gap (denoted by $E_g$) has a different behavior. When $L<6.2$~{\rm \AA}, $E_g$ is negative, manifesting a semi-metallic nature. As $L$ increases, $E_g$ first increases to a local maximum ($\sim $80~meV) at $L=6.4$~{\rm \AA}, and then decreases to zero at $L_{\rm c}$, forming a Dirac cone at $X$ point as shown in the inset of Fig.~\ref{bulk}(b). The above strain-dependent band structure analysis demonstrates that monolayer SnTe without strain has an inverted gap at $X$ point with a negative value.
According to the parity criteria \cite{fukane}, it can not be Z2 topological insulator as there are two $X$ points in the whole BZ. However, bulk SnTe has rocksalt crystal structure, and therefore its (001) monolayer thin films possess $z\rightarrow -z$ mirror symmetry which allows us to define the corresponding mirror Chern number $N_m$\cite{teofukane}. Following the similar analysis and argument \cite{jw_tf}, we can infer monolayer SnTe is a 2D TCI with non-zero mirror Chern number $|N_m|=2$.

The physical manifestation of non-zero mirror Chern number is the existence of gapless helical edge sates which are protected by $z\rightarrow -z$ mirror symmetry. By performing DFT calculations on the 20nm-wide ribbon of monolayer SnTe, we explicitly simulated the topological protected edge states.  As shown in Fig.~\ref{edge}, for the ribbon with small lattice constant 6.4~{\rm \AA} in the inverted regime (Fig.~\ref{bulk}(b)), there are two bands with linear dispersion crossing each other and connecting from the conduction bands to valence bands. Due to time-reversal symmetry, another pair of Dirac-type edge states is also present in the gap as shown in the inset of Fig.~\ref{edge}, which is consistent with the previous analysis $N_m=2$. Since the bands forming the Dirac point have different eigenvalues of $z\rightarrow -z$ mirror operation, both inter- and intra- back scattering are forbidden between those states. Thus, if the mirror symmetry is strictly preserved, those edge states can realize quantized transport.  Moreover, the mirror is connected with the spin in the system, and transport current should be also spin-polarized.

\begin{figure}[tbp]
\includegraphics[width=8.5cm]{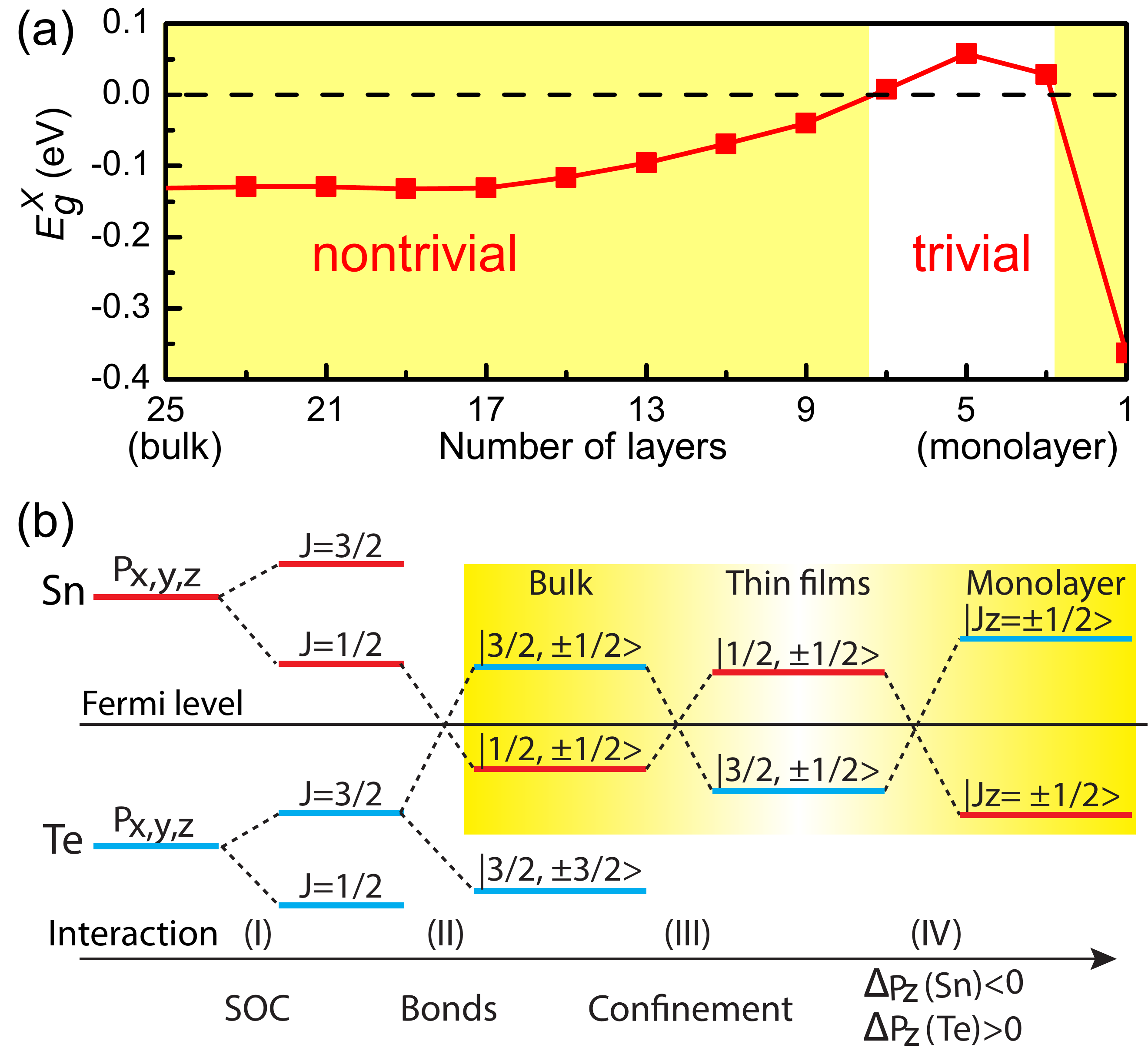}
\caption{(a)Thickness-dependent band gap at $X$ point of SnTe thin films. Except thin films with three, five, and seven atomic layers, all other thin films have {\it negative} inverted gap and thus nontrivial topological phases $|N_M|=2$.
(b) Schematic of band evolution in SnTe. (i) in the atomic limit, spin-orbital coupling splits the $p$ orbital; (ii) in the bulk limit ($z$ is set along [111] direction), strong chemical bonding between Sn and Te induces band inversion between $|J=3/2, J_z=\pm1/2\rangle$ of Sn and $|J=1/2, J_z=\pm1/2\rangle$ of Te, which drives bulk SnTe into 3D nontrivial TCI phase; (iii) for SnTe ultrathin films of 3, 5, and 7 layers, the confinement effect surpasses the inverted bulk gap and removes the band inversion formed in Step (ii); (iv) for monolayer ($z$ is set along [001]  direction), very different chemical bonding environment of $p_{x,y}$ and $p_z$ leads to the asymmetric crystal field effect. As a result, the onsite energy of $p_z$ will be lower (higher) than $p_{x,y}$ for Sn (Te), pushing all the $|J_z=\pm1/2\rangle$ states from Sn (Te) to lower (higher) energy and hence re-introducing band inversion.}
\label{orbit}
\end{figure}

We also performed DFT calculations for multi-layer thin films using the experimental lattice constant of bulk SnTe, $L_{\rm exp}=6.312$~{\rm \AA}, complementing our previous work~\cite{jw_tf}. Figure~\ref{orbit}(a) reveals a  {\it non-monotonically} change in ${E_g^{X}}$ with increasing thickness. Two critical thickness values divide the phase diagram into three parts. More specifically, except three, five, and seven layers of SnTe thin films, all other thin films have negative (or, inverted) ${E_g^{X}}$. For thin films of more than 17 layers, ${E_g^{X}}$ almost stays constant, reaching the gap of bulk SnTe. As the thickness gradually decreases, the confinement effect increases, driving ${E_g^{X}}$ to decrease and reach zero at about 7 layers. However, the gap didn't continue its monotonic change for thin films of less than 5 layers. Instead, gap closes again between monolayer and 3 layers due to the aforementioned negative inverted gap ${E_g^{X}}$ in monolayer SnTe. The thickness-dependent inverted gap is consistent with our previous work \cite{jw_tf} in the multi-layer case, providing a complete phase diagram shown in Fig.~\ref{orbit}(a).

To elucidate the underlying mechanism of band inversions, we analyzed the orbital evolution in detail and the result is presented in Fig.~\ref{orbit}(b). Previous works show that the fundamental gap of IV-VI semiconductor is around $L$ ($X$) points for bulk (thin films), and the bands near the Fermi level are mainly composed of $p$ orbitals of cation and anion (Sn and Te for SnTe) \cite{hsieh,jw_tf}. As shown in Fig.~\ref{orbit}(b), we start from the atomic limit where the chemical bonding between different atoms are ignored and focus on the $p$ orbitals. After taking SOC into account, the six-fold degenerate $p$ orbitals split into four-fold degenerate $J=3/2$ states and two-fold degenerate $J=1/2$ states. $J=3/2$ states are pushed up while $J=1/2$ states are lowered down in energy as shown in Step (i). We then included the hopping between those atomic orbitals (chemical bonds). For the bulk states, the $pp$ hopping between cation and anion vanishes at $L$ points, and the $sp$ and $pd$ hopping will play a vital role and lead to the four-fold degenerate $J=3/2$ states to further split into $J_z=\pm3/2$ and $J_z=\pm1/2$ states ($z$ along [111] direction). More importantly, the splitting is even strong enough to lower Sn's $|J=1/2,J_z=\pm1/2\rangle$ states than Te's $|J=3/2,J_z=\pm1/2\rangle$ states, driving SnTe into 3D TCI phase (Step (ii)). While for PbTe, the splitting induced by chemical bonding is not large enough to promote the band inversion although SOC is even stronger.

For the thin films, the states around $X$ points mainly inherit from the bulk states at $L$ point. When films are thick enough, all the properties of those states such as orbital weights and band orders are preserved, which ensure all the thin films thicker than a critical thickness (7 layers for SnTe) have negative inverted gap at $X$ points and possess the nontrivial topological phase, {\it i.e.} $|N_m|=2$. However, as the thickness decreases, the confinement effect will increase, hence enlarge the energy gap between anion and cation like in the normal semiconductor, giving rise to vanishing band inversion (see Step (iii) in Fig.~\ref{orbit}(b)) and topological phase transition as shown around 7 layers in Fig.~\ref{orbit}(a).

So far, all $p_{x,y,z}$ orbitals are treated equally. This is reasonable for bulk and thick films, since the difference between all $p_{x,y,z}$ orbitals is negligible. However, it is no longer true for {\it atomically} thin 2D materials as the chemical environment (crystal field) for $p_z$ and $p_{x,y}$ is completely different, where $z$ is along [001] direction. For $p_{x,y}$ orbitals, the crystal field is similar to the bulk where the $p_{x,y}$ orbitals between anion and cation form both $\sigma$ and $\pi$ type bonding. However, $\sigma$ bond is absent for the $p_z$ orbitals in monolayer. Such kind of asymmetric crystal field leads to the on-site energy of $p_z$ higher or lower than $p_{x,y}$ for cation (Sn) or anion (Te), respectively, {\it i.e.} $\rm{\Delta p_z(Sn)<0}$ and $\rm{\Delta p_z(Te)>0}$. This onsite energy difference $\rm{\Delta p_z}$ will push down (up) all the $J_z=\pm1/2$ states composed of Sn cation (Te anion) orbitals since all those states couple with each other and involve with $L_z=0$ states ({\it i.e.} $p_z$ orbital). From the DFT calculations, we found the asymmetric crystal field effect is strong enough to invert the bands between anion and cation again as shown in Step (iv) in Fig.~\ref{orbit}(b).
We also confirmed this band inversion mechanism and the corresponding topological phase transition by tuning the onsite energy difference in TB calculations.

\begin{figure}[tbp]
\includegraphics[width=8cm]{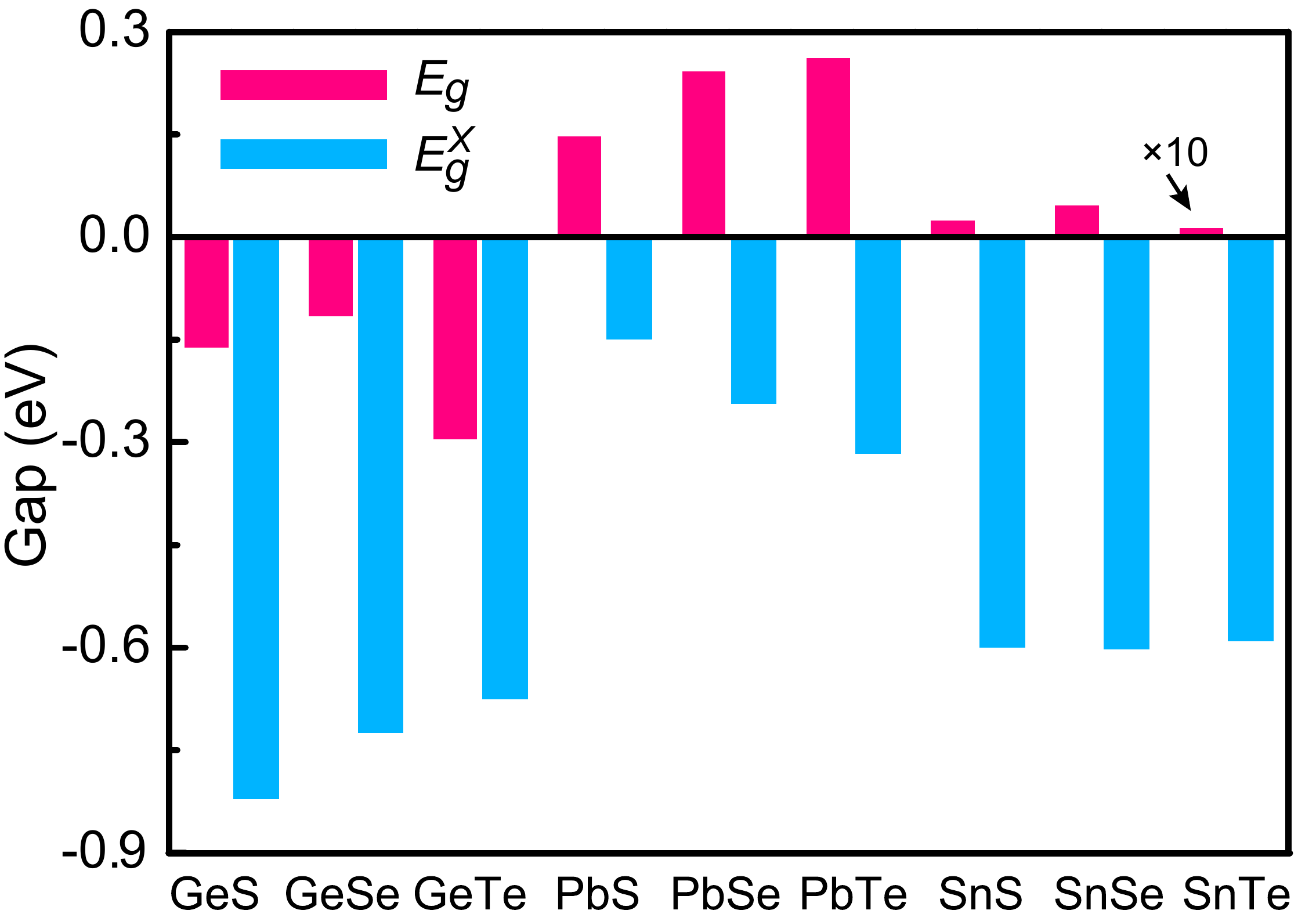}
\caption{The band gap at $X$ point (${E_g^{X}}$) and the fundamental gap $\rm{E_g}$ for all monolayer IV-VI semiconductors. All the materials have {\it negative} inverted gap at $X$ point and $|N_M|=2$. GeY (Y=S, Se, Te) are semi-metal and all other materials are fully gapped semiconductors. Among them, monolayer PbTe has the largest fundamental gap of 260 meV.}
\label{gap}
\end{figure}

It is worth to emphasize that the above band inversion mechanism in {\it atomically thin films} doesn't depend on the band inversion in {\it bulk}. Instead, it critically relies on the competition between the trivial band gap and the strong crystal field effect $\rm{\Delta p_z}$ on the $p_{x,y,z}$ orbitals in the monolayer. If $\rm{\Delta p_z}$ is big enough, we can also achieve nontrivial topological phases in monolayers of other materials even if their bulk forms don't have inverted gap. Motivated by this mechanism, we calculated other monolayer IV-VI semiconductors XY (X=Ge, Pb Sn and Y=S, Se, Te) and found that all them are 2D TCIs with $|N_m|=2$ in the optimized structure. The detailed band structures can be found in the Supplementary Materials. In Fig.~\ref{gap}, we plot the inverted gap at $X$ point (${E_g^{X}}$) and the fundamental gap $E_g$. Except GeY (Y=S, Se, Te) which are semi-metals\cite{note} with the largest inverted gap at $X$, all other materials including PbY and SnY (Y=S, Se, Te) are fully gapped semiconductors. Among the latter ones, monolayer PbTe has the largest fundamental gap of about 260 meV,  comparable with all existing and proposed 2D topological materials. We have confirmed nontrivial topological phases in all the materials by direct edge state calculations \cite{qian_tb, qian_green}. From the materials' perspective, PbTe is perhaps the most promising material to realize the novel application of topologically-protected edge states. First, PbTe holds the  highest mobility among all those materials \cite{ballistic}. Second, PbTe can have both {\it n}-type and {\it p}-type carriers \cite{wangna,yan} and its quality is easier to control in experiments \cite{yan}. In addition, except GeS, all other materials can realize the rocksalt structure under proper conditions \cite{Mariano, GeSe, GeTe}. Therefore, we expect that monolayer IV-VI semiconductors in rocksalt structure can be synthesized in experiments and realize the nontrivial topological phase presented above.

\section{CONCLUSION}

In conclusion, using first-principles calculations we theoretically studied the electric structures of monolayer IV-VI semiconductors along [001] direction. Remarkably, contrary to the expectation from the quantum confinement effect, all these materials are 2D TCIs with mirror Chern number $|N_m|=2$, and monolayer PbTe possesses the largest fundamental band gap of 260 meV. The 1D metallic edge states in these monolayer TCIs are protected by top-to-bottom mirror symmetry and robust against any perturbation preserving the mirror symmetry. The unexpected nontrivial topological phase originates from the strong asymmetric crystal field effect on $p_{x,y}$ and $p_{z}$ orbitals due to the unique characteristic of ultrathin 2D materials. Our work makes monolayer IV-IV semiconductors a promising materials platform to realize novel applications of topological edge states, and offers a new strategy to find other 2D topologically nontrivial materials.

\section{METHODS}

The calculations were performed in the framework of first-principles density-functional theory (DFT) implemented in the Vienna \emph{ab initio} simulation package (VASP) \cite{VASP}. We used the generalized gradient approximation (GGA) of exchange-correlation function in the Perdew-Burke-Ernzerhof (PBE) form \cite{PBE}. The projector augmented wave method \cite{PAW1} was applied to model the core electrons. Monkhorst-Pack \emph{\bf k}-point sampling of 10$\times$10$\times$1 was used for slab calculations. Energy cutoff of the planewave basis was fully tested for all the materials, and atomic structures were optimized with maximal residual forces smaller than 0.01 eV/{\AA}. Spin-orbit coupling (SOC) was included in all calculations.

\section{acknowledgement}
We thank Tim Hsieh for helpful discussions. This work was supported by the STC Center for Integrated Quantum Materials, NSF Grant No. DMR-1231319. X.Q. acknowledges the start-up funds from Texas A\&M University and the computational resources provided by Texas A\&M Supercomputing Facility.

{\it Note added:} During the preparation of our manuscript~\cite{march}, we learned of an independent work on the prediction of monolayer PbSe as TCI~\cite{Wrasse}.

\bibliographystyle{apsrev}

\begin{thebibliography}{10}
\bibitem{note}
{Topological invariant is still well-defined for semimetals although the fundamental gap is negative. It is because all direct gaps are non-zero in the whole Brillouin zone, separating the bands into the lower manifold and the higher manifold. Therefore, all the states in the lower bands form a closed manifold and allow us to define the corresponding mirror Chern number. From this sense and for the sake of the simplicity, we still name them as TCIs.}
\end{thebibliography}

\section{Supplementary Materials}

In the Supporting Information, we show the band structures of all monolayer IV-VI semiconductors XY (X=Ge, Sn, Pb and Y=S, Se, Te) in details.  All those materials have {\it negative} inverted gap at $X$ point, indicating that they are nontrivial 2D topological crystalline insulators with non-zero mirror Chern number $|N_m|=2$.  All the GeY (Y=S, Se, Te) are semi-metals, while all other materials are semiconductors. The large fundamental gap in PbY and SnY comes from the strong spin-orbit coupling of Pb and Sn.

We define the sign of inverted gap ${E_g^{X}}$ in all monolayer IV-VI semiconductors XY using the strain effect. We have also confirmed the nontrivial topological phases in all these materials by directly calculating the edge states.

\begin{figure*}[tbp]
\includegraphics[width=16cm]{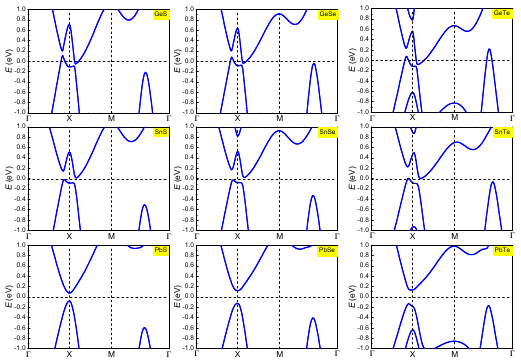}
\caption{Band structures of all monolayer IV-VI semiconductors XY (X=Ge, Sn, Pb and Y=S, Se, Te).}
\end{figure*}

\newpage

\end{document}